\documentclass[aps,onecolumn,12pt,nofootinbib,showpacs]{revtex4}
\usepackage[dvips]{graphicx} \usepackage{amssymb} \usepackage{psfrag}
\begin{document}
\begin{flushright}
\vspace*{-1.5cm}
BA-06-12\\
\end{flushright}
\vspace{-.1cm}
\title{Coleman-Weinberg Potential In Good Agreement With WMAP}
\author{Q. Shafi} \email{shafi@bartol.udel.edu}
\author{V. N. {\c S}eno$\breve{\textrm{g}}$uz} \email{nefer@udel.edu}
\affiliation{Bartol Research Institute, Department of Physics and Astronomy, University of Delaware, Newark, DE
19716, USA} 
\begin{abstract}
We briefly summarize and update a class of inflationary models from the early
eighties based on a quartic (Coleman-Weinberg) potential for a gauge
                      singlet scalar (inflaton) field.
For vacuum energy scales comparable to the grand unification scale, 
the scalar spectral index $n_s\simeq0.94$--$0.97$, in very good agreement with the
WMAP three year results. 
The tensor to scalar ratio $r\lesssim0.14$ while
$\alpha\equiv{\rm d}n_s/{\rm d}\ln k$ is $\simeq-10^{-3}$. 
An $SO(10)$ version naturally explains the observed baryon asymmetry via non-thermal leptogenesis.
\end{abstract}
\pacs{98.80.Cq}
\maketitle

An inflationary scenario \cite{Guth:1980zm,Linde:1981mu} 
may be termed successful if it satisfies the following criteria:

1) The total number of $e$-folds N during inflation is large enough to resolve the
horizon and flatness problems. Thus, $N\gtrsim$50--60, but
it can be somewhat smaller for low scale inflation.

2) The predictions are consistent with observations of the microwave
background and large scale structure formation. In particular,
the predictions for $n_s$, $r$ and $\alpha$ should be consistent with the
most recent WMAP results \cite{Spergel:2006hy} (see also \cite{Alabidi:2006qa} for
a brief survey of models).

3) Satisfactory resolution of the monopole problem in grand unified
theories (GUTs) is achieved.

4) Explanation of the origin of the observed baryon asymmetry is provided.

In this report we review a class of inflation models which appeared in the 
early eighties in the framework of non-supersymmetric GUTs and employed a GUT singlet scalar field
$\phi$ \cite{Shafi:1983bd,Pi:1984pv,Lazarides:1984pq}. 
These (Shafi-Vilenkin) models satisfy, as we will see, the above criteria and are based on a
Coleman-Weinberg (CW) potential \cite{Coleman:1973jx}
\begin{equation}
                  V(\phi)= V_0 + A \phi^4 \left[\ln\left(\frac{\phi^2}{M_*^2}\right) + C\right]
\end{equation}
where, following \cite{Shafi:1983bd} the renormalization mass  $M_* = 10^{ 18}$ GeV and
$V_0 ^{1/4}$ will specify the vacuum energy. The value of $C$ is fixed to cancel the cosmological
constant at the minimum. It is convenient to choose a physically
equivalent parametrization for $V(\phi)$ \cite{Albrecht:1984qt,Linde:2005ht}, namely
\begin{equation} \label{pot}
                  V(\phi)= A \phi^4 \left[\ln\left( \frac{\phi}{M}\right) -\frac{1}{4}\right] + \frac{A M^4}{4}\,,
\end{equation}
where $M$ denotes the $\phi$ VEV at the minimum. Note that $V(\phi=M)=0$,
and the vacuum energy density at the origin is given by
                   $V_0 = A M^4 /4$. 

The potential above is typical for the new inflation scenario \cite{Linde:1981mu},
where inflation takes place near the maximum. 
However, as we discuss below, 
depending on the value of $V_0$, the inflaton 
can have small or large values compared to the Planck scale during observable inflation.
In the latter case observable inflation takes place near the minimum and the
model mimics chaotic inflation \cite{Linde:1983gd}.

The original new inflation models attempted to explain the initial value of the inflaton through high-temperature
corrections to the potential. This mechanism does not work unless the inflaton is somewhat small
compared to the Planck scale at the Planck epoch \cite{Linde:2005ht}. However, the initial value of the 
inflaton could also be suppressed by a pre-inflationary phase. Here we will simply assume that the 
initial value of the inflaton is sufficiently small to allow enough $e$-folds.

The slow-roll parameters may be defined as \cite{Liddle:1992wi}
\begin{equation}
\epsilon =
 \frac{1}{2}\left(\frac{V'}{V}\right)^2 \,,\quad
\eta=\left(\frac{V''}{V}\right) \,,\quad
\xi^2=\left(\frac{V'\, V'''}{V^2}\right) \,.
\end{equation}
(Here and below we use units $m_P=1$, where $m_P\simeq2.4\times10^{18}$ GeV
is the reduced Planck mass.) 
The slow-roll approximation is valid if the slow-roll conditions
$\epsilon \ll 1$ and $\eta \ll 1$ hold.
In this case the spectral index
$n_\mathrm{s}$, the tensor-to-scalar ratio
$r$ and the running of the spectral index
$\alpha\equiv\mathrm{d} n_\mathrm{s}/\mathrm{d} \ln k$ are given by
\begin{eqnarray}
n_\mathrm{s} \!&\simeq&\! 1 - 6 \epsilon + 2 \eta \label{ns}\\
r \!&\simeq&\! 16 \epsilon \\
\alpha \!\!&\simeq&\!\!
16 \epsilon \eta - 24 \epsilon^2 - 2 \xi^2.
\end{eqnarray}

The number of $e$-folds after the comoving scale $l_0=2\pi/k_0$ has crossed the horizon is
given by
\begin{equation} \label{efold1}
N_0=\frac{1}{2}\int^{\phi_0}_{\phi_e}\frac{H(\phi)\rm{d}\phi}{H'(\phi)} \end{equation}
where $\phi_0$ is the value of the field when the scale corresponding to $k_0$
exits the horizon and $\phi_e$ is the value of the field at the end of inflation.
This value is given by the condition $2(H'(\phi)/H(\phi))^2=1$, which can be calculated
from the Hamilton-Jacobi equation \cite{Salopek:1990jq}
\begin{equation}
[H'(\phi)]^2-\frac{3}{2}H^2(\phi)=-\frac{1}{2}V(\phi)\,.
\end{equation}
The amplitude of the curvature perturbation $\mathcal{P^{{\rm 1/2}}_R}$ is given by
\begin{equation} \label{perturb}
\mathcal{P^{{\rm 1/2}}_R}=\frac{1}{2\sqrt{3}\pi }\frac{V^{3/2}}{|V'|}\,.
\end{equation}
To calculate the magnitude of $A$ and the inflationary parameters, we use these standard
equations above. We also include the first order corrections in the slow roll expansion
for $\mathcal{P^{{\rm 1/2}}_R}$ and the spectral index $n_s$ \cite{Stewart:1993bc}.\footnote{The 
fractional error in $\mathcal{P^{{\rm 1/2}}_R}$ from the slow
roll approximation is of order $\epsilon$ and $\eta$ (assuming these parameters remain $\ll1$).
This leads to an error in $n_s$ of order $\xi^2$, which is $\sim10^{-3}$ in the present model. 
Comparing to the WMAP errors, this precision seems quite adequate.
However, in anticipation of the Planck mission, it may be desirable
to consider improvements.}
The WMAP value for $\mathcal{P^{{\rm 1/2}}_R}$ is $4.86\times10^{-5}$ for $k_0 = 0.002$ Mpc$^{-1}$.
$N_0$ corresponding to the same scale is $\simeq53+(2/3)\ln(V(\phi_0) ^{1/4}/10^{15}\ \rm{GeV})+(1/3)\ln(T_r/10^{9}\ \rm{GeV})$.
(The expression for $N_0$ assumes a standard thermal history \cite{Dodelson:2003vq}. See \cite{Lyth:1998xn} for reviews.)
We assume reheating is efficient enough such that $T_r=m_{\phi}$, where the mass of the inflaton
$m_{\phi}=2\sqrt{A}M$. In practice, we expect $T_r$ to be somewhat below $m_{\phi}$ \cite{Shafi:1983bd}.

\begin{figure}[thb] 
\psfrag{1-n}{\scriptsize{$1-n_s$}}
\psfrag{r}{\scriptsize{$r$}}
\includegraphics[angle=0, width=8cm]{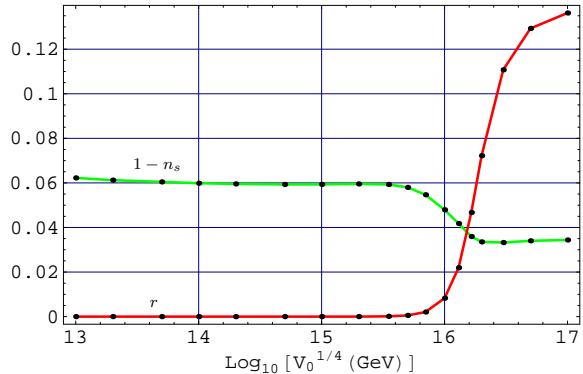} 
 \vspace{-.5cm} 
\caption{$1-n_s$ and $r$ vs. $\log[V^{1/4}_0{\,\rm(GeV)}]$ for the potential in Eq. (\ref{pot}).} \label{figgg}
\end{figure}

In Table I  and Fig. \ref{figgg} we display the predictions for $n_s$, $\alpha$ and $r$, with
the vacuum energy scale $V_0 ^{1/4}$ varying from $10^{13}$ GeV to  $10^{17}$ GeV.
The parameters have a slight dependence on the reheat temperature, as can be seen from
the expression for $N_0$. As an example, if we assume instant reheating ($T_r\simeq V(\phi_0)^{1/4}$), 
$n_s$ would increase to 0.941 and 0.943 for $V_0^{1/4} =10^{13}$ GeV and $V_0^{1/4} =10^{15}$ GeV respectively. 

\begin{table}[t] 
{\centering
\caption{The inflationary parameters for the Shafi-Vilenkin model with the potential in Eq. (\ref{pot}) 
($m_P=1$)} \label{tablo}
\resizebox{!}{3.72cm}{
\begin{tabular}{r@{\hspace{.5cm}}r@{\hspace{.5cm}}r@{\hspace{.5cm}}r@{\hspace{.5cm}}r@{\hspace{.5cm}}r@{\hspace{.5cm}}r@{\hspace{.5cm}}r@{\hspace{.5cm}}r}
\hline \hline
 $V^{1/4}_0$(GeV) & $A(10^{-14})$ & M & $\phi_e$ & $\phi_0$ & $V(\phi_0)^{1/4}$(GeV) & $n_s$ & $\alpha(-10^{-3})$ & $r$   \\
\hline
$10^{13}$ & 1.0 & 0.018 &  0.010 &  $3.0\times10^{-6}$  &$\approx V_0^{1/4}$ &0.938 & 1.4 & $9\times10^{-15}$  \\
\hline
$5\times10^{13}$ & 1.2 & 0.088 &  0.050  & $7.5\times10^{-5}$ &$\approx V_0^{1/4}$ &0.940 & 1.3 & $5\times10^{-12}$   \\
\hline
$10^{14}$ & 1.3 & 0.17 & 0.10  &  $3.0\times10^{-4}$ &$\approx V_0^{1/4}$ &0.940 & 1.2 & $9\times10^{-11}$ \\
\hline
$5\times10^{14}$ & 1.9 & 0.79 & 0.51   & $7.5\times10^{-3}$ & $\approx V_0^{1/4}$&0.941 & 1.2 & $5\times10^{-8}$ \\
\hline
$10^{15}$ & 2.3 & 1.5 & 1.1  & 0.030 &$\approx V_0^{1/4}$ &0.941 & 1.2 & $9\times10^{-7}$   \\
\hline
$5\times10^{15}$ & 4.8 & 6.2 & 5.1  &  0.71 & $\approx V_0^{1/4}$&0.942 & 1.0 & $5\times10^{-4}$ \\
\hline
$10^{16}$ & 5.2 & 12 & 10  &  3.2 & $9.9\times10^{15}$ & 0.952 & 1.0 & $8\times10^{-3}$ \\
\hline
$2\times10^{16}$ & 1.1 & 36 &35   & 23 &$1.7\times10^{16}$ &0.966 & 0.6 & $0.07$ \\
\hline
$3\times10^{16}$ & .17 & 86 &  85  & 72&$1.9\times10^{16}$ &0.967 & 0.6 & $0.11$ \\ 
\hline
$10^{17}$ & .001 & 1035 &  1034  & 1020&$2.0\times10^{16}$ &0.966 & 0.6 & $0.14$ \\
\hline \hline
\end{tabular} }
\par} \centering 
\end{table}

For $V_0^{1/4}\lesssim10^{16}$ GeV, the inflaton field remains smaller than the Planck scale, 
and the inflationary parameters are
similar to those for new inflation models with $V=V_0(1-(\phi/\mu)^4)$: $n_s\simeq1-(3/N_0)$,
$\alpha\simeq (n_s-1)/N_0$.
As the vacuum energy is lowered, $N_0$ becomes smaller and
$n_s$ deviates further from unity. However, $n_s$ remains within $2\sigma$ of the WMAP
best fit value (for negligible $r$) $0.951^{0.015}_{-0.019}$ \cite{Spergel:2006hy} 
even for $V_0^{1/4}$ as low as $10^5$ GeV. Inflation 
with CW potential at low scales is discussed in Ref. \cite{Knox:1992iy}.

For $V_0^{1/4}\gtrsim10^{16}$ GeV, the inflaton is larger than the Planck scale
during observable inflation. Observable inflation then occurs closer to the minimum 
where the potential is effectively $V=(1/2) m_{\phi}^2 \Delta\phi^2$,
$\Delta\phi=M-\phi$ denoting the deviation of the field from the minimum.
This well-known monomial model \cite{Linde:1983gd} predicts $m_{\phi}\simeq2\times10^{13}$ GeV and
$\Delta\phi_0\simeq2\sqrt{N_0}$, corresponding to $V(\phi_0)\simeq(2\times10^{16}$ GeV$)^4$. 
For the $\phi^2$ potential to be a good approximation, $V_0$ must be greater than 
this value. Then the inflationary parameters no longer depend on $V_0$ and
approach the predictions for the $\phi^2$ potential.

\begin{figure}[thb] 
\includegraphics[angle=0, width=8cm]{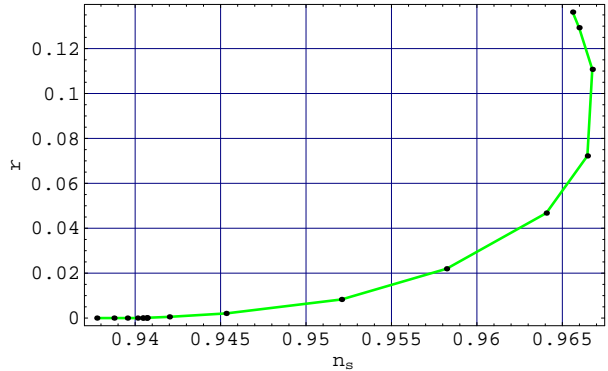} 
 \vspace{-.5cm} 
\caption{$r$ vs. $n_s$ for the potential in Eq. (\ref{pot}).} \label{fig:222}
\end{figure}

\begin{figure}[thb] 
\includegraphics[angle=0, width=8cm]{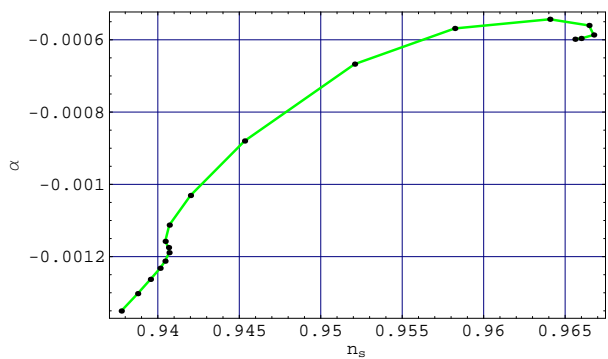} 
 \vspace{-.5cm} 
\caption{$\alpha$ vs. $n_s$ for the potential in Eq. (\ref{pot}).} \label{fig:2222}
\end{figure}

The tensor to scalar ratio $r$ versus the spectral index $n_s$ is displayed in Fig. \ref{fig:222}.
The values are in very good agreement with the recent WMAP results
(see Fig. 14 of Ref. \cite{Spergel:2006hy}). 
The running of the spectral index is negligible, as in most inflation models (Fig. \ref{fig:2222}).
Note that the WMAP data favor a large running spectral index. 
This is an important result if confirmed but currently has little statistical
significance.

In the context of non-supersymmetric GUTs, $V_0^{1/4}$ is related to the unification
scale, and is typically a factor of 3--4 smaller than the superheavy gauge boson masses
due to the loop factor in the CW potential. 
The unification scale for non-supersymmetric GUTs is typically $10^{14}$--$10^{15}$
GeV, although it is possible to have higher scales, for instance associating
inflation with $SO(10)$ breaking via $SU(5)$.

The reader may worry about proton decay with gauge boson masses of order
$10^{14}$--$10^{15}$ GeV.
In the $SU(5)$ model \cite{Georgi:1974sy}, in particular, a two-loop
renormalization group analysis of the standard model gauge couplings yields
masses for the superheavy gauge bosons of order $1\times10^{14}$--$5\times10^{14}$ GeV
\cite{Dorsner:2005ii}. This is consistent with the  SuperK proton lifetime limits \cite{Eidelman:2004wy},
provided one assumes strong flavor suppression of the relevant dimension six
gauge mediated proton decay coefficients. 
If no suppression is assumed the gauge boson masses should have masses
close to $10^{15}$ GeV or higher \cite{Ilia}.

For the Shafi-Vilenkin model
in $SU(5)$, the tree level scalar potential
contains the term $(1/2)\lambda\phi^2{\rm Tr}\Phi^2$ with $\Phi$ being the Higgs adjoint,
and $A\sim1.5\times10^{-2}\lambda^2$ \cite{Shafi:1983bd,Linde:2005ht}. Inflation requires 
$A\sim10^{-14}$, corresponding to $\lambda\sim10^{-6}$.

This model has been extended to $SO(10)$
in Ref. \cite{Lazarides:1984pq}. The breaking of $SO(10)$ to the standard model proceeds, for
example, via the subgroup $G_{422} = SU(4)_c \times SU(2)_L \times SU(2)_R$ \cite{Pati:1974yy}. A renormalization
group analysis shows that the symmetry breaking scale for $SO(10)$ is of order $10^{15}$
GeV, while $G_{422}$ breaks at an intermediate scale $M_I\sim 10^{12}$ GeV
\cite{Rajpoot:1980xy}. (This is
intriguingly close to the scale needed to resolve the strong CP problem and
produce cold dark matter axions.)
The predictions for $n_s$, $\alpha$ and $r$ are essentially identical
to the SU(5) case. There is one amusing consequence though which may be worth
mentioning here. The monopoles associated with the breaking of $SO(10)$ to $G_{422}$
are inflated away. However, the breaking of $G_{422}$ to
the SM gauge symmetry yields doubly charged monopoles \cite{Lazarides:1980cc}, whose mass is
of order $10^{13}$ GeV. These may be present in our galaxy  at a flux level of
$10^{-16}\ {\rm cm}^{-2}$ s$^{-1}$ sr$^{-1}$ \cite{Lazarides:1984pq}.

As stated earlier, before an inflationary model can be deemed successful, it must contain
a mechanism for generating the observed baryon asymmetry in the universe. In
the $SU(5)$ case the color higgs triplets produced by the
inflaton decay can generate the baryon asymmetry, provided  the higgs sector of
the model has the required amount of CP violation \cite{Shafi:1983bd}.

The discovery of neutrino oscillations requires that we introduce SU(5)
singlet right handed neutrinos, presumably three of them, to implement
the seesaw mechanism and generate the desired masses for the light
neutrinos. In this case it is natural to generate the observed baryon
asymmetry via leptogenesis \cite{Fukugita:1986hr} (for non-thermal leptogenesis see Ref. \cite{Lazarides:1991wu})
by introducing the couplings
                   $N_i N_j \phi^2 / m_P$,
where $N_i$ (i=1,2,3) denote the right handed neutrinos, and the renormalizable
coupling to $\phi$ is absent because of the assumed discrete
symmetry. By suitably adjusting the Yukawa coefficients one can arrange
that the $\phi$ field decays into the right handed neutrinos.
Note that the presence of the above Yukawa couplings then allows one
to make the color triplets heavier, of order $10^{14}$ GeV, thereby avoiding
any potential conflict with proton decay. In the $SO(10)$ model,
leptogenesis is almost automatic \cite{Lazarides:1984pq}.

Finally, it is worth noting that new inflation models have also been 
considered in the framework of supersymmetric GUTs, taking account of
supergravity corrections. In Ref. \cite{Senoguz:2004ky}, for instance, it is
shown that the spectral index $n_s$ is less than 0.98, with values between
0.94 and 0.96  plausible.  Furthermore, reheat temperatures as low
as $10^4$--$10^6$ GeV can be realized to satisfy the gravitino constraint.
In these models the tensor to scalar ratio $r$ is
tiny, of order $10^{-3}$ or less, and $\alpha\sim-10^{-3}$. 

To summarize, we have briefly reviewed and updated a class of realistic
inflation models based on a quartic CW potential for a gauge singlet
inflaton field. We find very good agreement between the model
predictions and the three year WMAP data. An interesting feature
is the observation that if the vacuum energy that drives inflation
exceeds $10^{16}$ GeV, the inflaton during observable inflation exceeds
the Planck mass in value. As a consequence, there is transition
from new to chaotic inflation (or more precisely the monomial potential model), 
and the scalar spectral index and $r$ acquire limiting values of $0.966$ and $0.14$ respectively.

\subsection*{Acknowledgements}
This work is partially supported by the US Department of Energy under contract number DE-FG02-91ER40626 (Q.S. and V.N.S.),
and by a University of Delaware graduate fellowship (V.N.S.).

\end{document}